\documentclass[twocolumn,showpacs,preprintnumbers,amsmath,amssymb]{revtex4}

\usepackage{graphicx}
\usepackage{dcolumn}
\usepackage{bm}

\newcommand{\bq}{\begin{equation}}
\newcommand{\eq}{\end{equation}}
\newcommand{\bqa}{\begin{eqnarray}}
\newcommand{\eqa}{\end{eqnarray}}
\newcommand{\nn}{\nonumber \\}
\def\ij {\langle i j \rangle}

\def\be     {\begin{equation}}
\def\ee     {\end{equation}}
\def\bea        {\begin{eqnarray}}
\def\eea        {\end{eqnarray}}
\def\bnn    {\begin{eqnarray*}}
\def\enn    {\end{eqnarray*}}

\begin{document}

\title{Spin-gapped incoherent metal with preformed pairing in the doped antiferromagnetic Mott insulator}
\author{Ki-Seok Kim and Mun Dae Kim}
\affiliation{ School of Physics, Korea Institute for Advanced
Study, Seoul 130-012, Korea }
\date{\today}

\begin{abstract}
We investigate how the antiferromagnetic Mott insulator evolves
into the d-wave BCS superconductor through hole doping. Allowing
spin fluctuations in the strong coupling approach, we find a
spin-gapped incoherent metal with preformed pairing as an
intermediate phase between the antiferromagnetic Mott insulator
and d-wave superconductor. This non-Fermi liquid metal is
identified with an infrared stable fixed point in the
spin-decomposition gauge theory, analogous to the spin liquid
insulator in the slave-boson gauge theory. We consider the single
particle spectrum and dynamical spin susceptibility in the
anomalous metallic phase, and discuss physical implications.
\end{abstract}


\maketitle

\section{Introduction}

The problem of doped Mott insulators has been one of the central
interests in modern condensed matter physics. In particular, the
route or mechanism from the parent Mott insulating state to the
superconducting phase lies at the heart of the research in
strongly correlated electrons. Such a route  depends
on the nature of the parent Mott insulating phase. Generally
speaking, the Mott insulator can be characterized based on its
symmetry breaking patterns.\cite{Sachdev_Review_MI} Usually, it
exhibits symmetry breaking associated with spin rotations or
lattice translations, thus causing possible long-range orders.
However, when frustration effects are strong enough to kill such
orders, the resulting Mott insulator is symmetric, called a spin
liquid Mott insulator.\cite{RMP} Doping to the symmetry-broken and
symmetric Mott insulators would result in different routes to
superconductivity.

There exist analytical frameworks appropriate to each doped Mott
insulator. The doped spin liquid Mott insulator can be described
by the slave-boson representation of the t-J model\cite{RMP} while
the doped antiferromagnetic Mott insulator can be captured by its
slave-fermion description.\cite{Shankar,Reviews,Slave_Fermion} One
of the translationally symmetry-broken insulators may be described
by the bond-operator formalism.\cite{Bond_Operator} Our main
interest in this paper is doping to the antiferromagnetic Mott
insulator, and seeing the emergence of superconductivity from the
doped antiferromagnetic Mott insulator. In this respect it seems
natural to adopt the slave-fermion representation. Unfortunately,
it is believed that the slave-fermion framework does not give rise to
the superconductivity naturally. In the next section we review the
slave-fermion representation of the t-J model, comparing with the
slave-boson approach and discuss the reason why it is not easy to
obtain superconductivity in the slave-fermion description. In
this study we employ the CP$^1$ spin-decomposition
approach\cite{Auerbach} which allows superconductivity. The
CP$^1$ representation follows the same philosophy as the
slave-fermion description in the physical point of view.

We have two symmetry breaking phases at both sides of doping, that
is, antiferromagnetic Mott insulator at half filling and d-wave
superconductor at large doping. The antiferromagnetic Mott
insulator is described by the O(3) nonlinear $\sigma$ model, an
effective field theory of the antiferromagnetic Heisenberg model
for spin dynamics. Here charge dynamics is almost frozen, thus
safely ignored in the low energy limit.\cite{Auerbach} On the
other hand, the superconducting phase follows the BCS-type
approach with d-wave pairing. We show that the CP$^1$
decomposition approach recovers such known theoretical limits
naturally.

The main object of this paper is to find the route connecting
these well-known symmetry-breaking phases. When holes are doped
into the antiferromagnetic Mott insulator, charge fluctuations
would be gapless. Then, metallic physics may arise in
the absence of disorder, if superconductivity is not taken into
account. Actually, numerical simulations based on the t-J model
have shown metallic properties.\cite{Simulation} An interesting
question is whether such an intermediate metallic phase will
survive or not when superconductivity is allowed. The metallic
phase may be unstable to disappear. Then, there will be a
coexisting phase of antiferromagnetism and superconductivity or
the first order transition between them. However, in this paper we
show that the metallic state exists indeed as an intermediate
phase between the symmetry breaking phases. This kind of metallic
phase has been recently argued to appear in the slave-fermion
framework.\cite{Slave_Fermion} In the CP$^1$ framework we find
that the metallic phase turns into a d-wave superconducting state
as holes are doped further. We discuss the nature of this metallic
phase and find that such a phase is identified with a non-Fermi
liquid metal.

Starting from the BCS-HF (Hartree-Fock) model [Eq.
(\ref{BCS-HF})], we derive an effective theory: the CP$^1$
representation of the O(3) nonlinear $\sigma$ model for spin
dynamics and the BCS-HF theory for charge dynamics, coupled via
U(1) spin-gauge fluctuations [Eq. (\ref{SRGT})].
Here CP$^1$ spin-gauge fluctuations play the role
of pairing fluctuations in the pairing term of the fermion sector.
As a result, we find a spin-gapped incoherent metal with preformed
pairing excitations as an intermediate phase between the
antiferromagnetic Mott insulator and d-wave superconductor. In the
last section we argue that the spin-gapped incoherent metal with
preformed pairing is analogous to the spin-gapped
"superconducting" state in one dimension.\cite{Shankar} The
present study not only generalizes the one dimensional work of
Shankar\cite{Shankar} into two dimensions, but also extends the
previous studies\cite{Reviews} into an incoherent regime where
spin-boson excitations are gapped.

\section{Review of the slave-fermion approach}

To clarify the connection between the present approach and
slave-fermion description, it is necessary to review the
slave-fermion representation of the t-J model. In addition, to
understand the reason why superconductivity does not arise
naturally in the slave-fermion approach, we compare the
slave-fermion representation of the t-J model with the slave-boson
framework. Consider the t-J Hamiltonian
\bqa && H = -t\sum_{\ij\sigma}(c_{i\sigma}^{\dagger}c_{j\sigma} + H.c.) + J
\sum_{\ij}(\vec{S}_{i}\cdot\vec{S}_{j} - \frac{1}{4}n_{i}n_{j})
\nn \eqa with the constraint
$\sum_{\sigma}c_{i\sigma}^{\dagger}c_{i\sigma} \leq 1$. This
inequality constraint is not easy to handle. In order to treat
such a constraint a slave-particle decomposition approach can be
introduced, since it turns the inequality constraint into an
equality one. Depending on the statistics of charge and spin
degrees of freedom, one can decompose the electron operator
according to the following ways, \bqa && c_{i\sigma} =
b_{i}^{\dagger}f_{i\sigma} , ~~~~~ b_{i}^{\dagger}b_{i} +
\sum_{\sigma}f_{i\sigma}^{\dagger}f_{i\sigma} = 1 , \nn &&
c_{i\sigma} = \psi_{i}^{\dagger}b_{i\sigma} , ~~~~~
\psi_{i}^{\dagger}\psi_{i} +
\sum_{\sigma}b_{i\sigma}^{\dagger}b_{i\sigma} = 1 . \eqa Here the
first line shows the slave-boson representation with the bosonic
charge and fermionic spin, and the second line the slave-fermion
one with the fermionic charge and bosonic spin.

In each representation the Heisenberg term can be expressed as
\bqa && J \sum_{\ij}(\vec{S}_{i}\cdot\vec{S}_{j} -
\frac{1}{4}n_{i}n_{j}) = -
\frac{J}{2}\sum_{\ij}\hat{\Delta}_{ij}^{\dagger}\hat{\Delta}_{ij},
\nn && \mbox{ slave-boson :   } \hat{\Delta}_{ij}
=\sum_{\sigma\sigma'}\epsilon_{\sigma\sigma'}
f_{i\sigma}f_{j\sigma'},   \nn && \mbox{ slave-fermion :   }
\hat{\Delta}_{ij} = \sum_{\sigma\sigma'}\epsilon_{\sigma\sigma'}
b_{i\sigma}b_{j\sigma'}   \eqa for the pairing channel. Inserting
each decomposition representation into the t-J model with Eq. (3),
and performing the Hubbard-Stratonovich transformation for the
exchange hopping and Heisenberg pairing channels, one can find
each effective Lagrangian \bqa && L_{SB} =
\sum_{i}b_{i}^{\dagger}\partial_{\tau}b_{i} -
t\sum_{\ij}(b_{i}^{\dagger}\chi_{ij}^{b}b_{j} + H.c.) \nn && +
\sum_{i\sigma}f_{i\sigma}^{\dagger}(\partial_{\tau} -
\mu)f_{i\sigma} - t
\sum_{\ij\sigma}(f_{i\sigma}^{\dagger}\chi_{ij}^{f}f_{j\sigma} +
H.c.) \nn && -
\sum_{\ij\sigma\sigma'}(\Delta_{ij}^{\dagger}\epsilon_{\sigma\sigma'}
f_{i\sigma}f_{j\sigma'} + H.c.) \nn && + i
\sum_{i}\lambda_{i}(b_{i}^{\dagger}b_{i} +
\sum_{\sigma}f_{i\sigma}^{\dagger}f_{i\sigma} - 1) \nn && + t
\sum_{i}(\chi_{ij}^{b}\chi_{ij}^{f} + h.c.) +
\frac{1}{2J}\sum_{\ij}|\Delta_{ij}|^{2} , \nn && L_{SF} =
\sum_{i}\psi_{i}^{\dagger}\partial_{\tau}\psi_{i} +
t\sum_{\ij}(\psi_{i}^{\dagger}\chi_{ij}^{\psi}\psi_{j} + H.c.) \nn
&& + \sum_{i\sigma}b_{i\sigma}^{\dagger}(\partial_{\tau} -
\mu)b_{i\sigma} - t
\sum_{\ij\sigma}(b_{i\sigma}^{\dagger}\chi_{ij}^{b}b_{j\sigma} +
H.c.) \nn && -
\sum_{\ij\sigma\sigma'}(\Delta_{ij}^{\dagger}\epsilon_{\sigma\sigma'}
b_{i\sigma}b_{j\sigma'} + H.c.) \nn && + i
\sum_{i}\lambda_{i}(\psi_{i}^{\dagger}\psi_{i} +
\sum_{\sigma}b_{i\sigma}^{\dagger}b_{i\sigma} - 1) \nn && + t
\sum_{i}(\chi_{ij}^{\psi}\chi_{ij}^{b} + h.c.) +
\frac{1}{2J}\sum_{\ij}|\Delta_{ij}|^{2} \eqa in the slave-boson
and slave-fermion representations, respectively. Here,
$\chi_{ij}^{b,f,\psi}$ is the effective hopping parameter and
$\Delta_{ij}$ is the pairing order parameter, both of which will
be determined self-consistently. $\lambda_{i}$ is a Lagrange
multiplier field to impose single occupancy constraint
associated with each decomposition.

Our interest lies in the region where antiferromagnetic
correlations are enhanced, captured by spin-singlet pairing
fluctuations ($\Delta_{ij} \not= 0$). When charge fluctuations are
frozen at half filling, the slave-boson approach gives rise to a
spin liquid Mott insulating phase, which is a starting point in
the slave-boson context. As holes are doped into the spin liquid
phase, bosonic charge degrees of freedom becomes condensed to form
electronic Cooper pairs, resulting in superconductivity. In this
scenario the finite-temperature pseudogap physics is governed by
the spin liquid physics for spin fluctuations. On the other hand,
the slave-fermion approach will give rise to an antiferromagnetic
long-range order at half filling via condensation of bosonic spin
degrees of freedom, where fermionic charge fluctuations are gapped
to be frozen. As holes are doped into the antiferromagnetic Mott
insulator, bosonic spin fluctuations can be gapped and fermionic
charge excitations will become gapless to show a metallic
phase.\cite{Shankar,Slave_Fermion,Comment_SF} Such a spin-gapped
metallic state is expected to govern the pseudogap physics in the
slave-fermion context.\cite{Slave_Fermion} As will be discussed
later, the mathematical structures of both theories are nearly
identical. The spin liquid Mott insulating phase can be identified
with an infrared stable fixed point of the effective fermion-gauge
Lagrangian with damped gauge fluctuations, where fermions carry
spin degrees of freedom. On the other hand, the spin-gapped
metallic phase would be characterized by the same fixed point of
nearly the same Lagrangian, but fermions carry charge degrees of
freedom here. Since we are  focusing on doping the antiferromagnetic Mott
insulator, the slave-fermion scheme is more appropriate.

However, there is one serious difficulty in the slave-fermion
approach. It is not easy to obtain superconductivity in the
slave-fermion context. In the slave-boson representation
condensation of bosonic charge fluctuations causes
superconductivity in the presence of spin-singlet pairing
excitations, while fermionic charge
degrees of freedom cannot be condensed in the slave-fermion representation.
One possible way is to
introduce pairing fluctuations of fermionic charge degrees of
freedom. Unfortunately, such pairing interactions between
fermionic charge degrees of freedom do not arise in the naive
mean-field approximation. The Heisenberg term can be written in
terms of only bosonic spin degrees of freedom, if the
slave-fermion constraint is used appropriately. One can show that
such pairing interactions can arise from gauge fluctuations,
originating from the slave-fermion decomposition. This is
certainly possible, but beyond the mean-field approximation. Our
objective is to construct an effective self-consistent mean-field
theory in the slave-fermion scheme. In addition, we are to examine
the fate of such an anomalous metallic phase when superconducting
instability is allowed. In this paper we show that the spin-gapped
metal appears as an intermediate phase between the
antiferromagnetic Mott insulator and d-wave superconductor.

\section{Formulation}

\subsection{Model Hamiltonian}

In the previous section we have discussed that the slave-fermion
approach allows an anomalous spin-gapped metal via doping to the
antiferromagnetic Mott insulator, and such a phase is analogous to
the spin liquid state in the slave-boson context. In addition, we
argued that it is nontrivial to find superconductivity in the
saddle-point analysis of the slave-fermion framework due to the
statistics of charge degrees of freedom and the absence of pairing
interactions in the mean-field level. In this respect it is
necessary to find another representation, keeping the
slave-fermion scheme. We consider the t-J-U model
Hamiltonian\cite{BCS_Hubbard1} \bqa &&  H = -
t\sum_{\ij\sigma}(c_{i\sigma}^{\dagger}c_{j\sigma} + H. c.) \nn &&
- J
\sum_{\ij}\sum_{\alpha\beta\gamma\delta}(\epsilon_{\alpha\beta}c_{i\alpha}^{\dagger}c_{j\beta}^{\dagger})
(\epsilon_{\gamma\delta}c_{i\gamma}c_{j\delta}) +
U\sum_{i}n_{i\uparrow}n_{i\downarrow} , \eqa where the Heisenberg
term is expressed as a pairing channel.

One may argue that interaction terms in this model Hamiltonian are
redundant since the J term can be generated from the t-U terms via
virtual hopping processes. Although this statement is basically
correct, actually the mean-field analysis can hardly captures  the
effects of exchange interactions if one starts from the Hubbard
model. One can understand such a model as
follows.\cite{BCS_Hubbard1} Starting from the t-U model and
integrating out high energy degrees of freedom, one would find the
exchange interaction term in the intermediate step of the
renormalization group analysis. If the on-site repulsion is not
infinitely large, one can keep such an interaction in the
intermediate level. Actually, the t-J-U model connects the two
limiting cases smoothly. In large-U limit this model is reduced to
the t-J model while it recovers the t-U model in small-J limit.
Remember that the slave-fermion approach in the large-U limit is
not appropriate for describing superconductivity.

Performing the Hubbard-Stratonovich transformation for the pairing
channel, we obtain the BCS-Hubbard Hamiltonian as an appropriate
model for the doped antiferromagnetic Mott
insulator\cite{BCS_Hubbard1,BCS_Hubbard2} \bqa && H = -
t\sum_{\ij\sigma}(c_{i\sigma}^{\dagger}c_{j\sigma} + H. c.) -
\sum_{\ij}\bigl[\Delta_{ij}(c_{i\uparrow}^{\dagger}c_{j\downarrow}^{\dagger}
- c_{i\downarrow}^{\dagger}c_{j\uparrow}^{\dagger}) \nn && + H.
c.\bigr] + \frac{1}{J}\sum_{\ij}|\Delta_{ij}|^{2} +
U\sum_{i}n_{i\uparrow}n_{i\downarrow} , \label{BCS-HF} \eqa where
the competing nature arising from the density-phase uncertainty is
introduced; the J term causes local pairing $(\Delta_{ij})$ of
electrons while the Hubbard-U term suppresses local charge
fluctuations, thus breaking phase coherence of electron pairs.

\begin{figure}[b]
\vspace*{7cm} \includegraphics{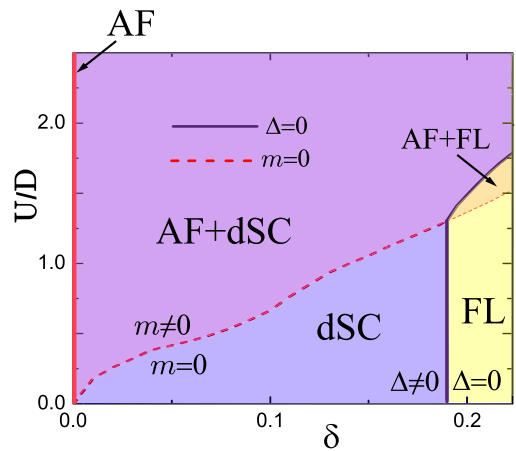} \vspace{-0.5cm}
\caption{(Color online) BCS-HF phase diagram with $J/D=0.1$. Separation between
the dSC and FL via the straight line is an artifact of the BCS-HF
analysis. See the text.} \label{Fig1}
\end{figure}

Decomposing the Hubbard-U term into the charge and spin channels,
we obtain the BCS-HF Lagrangian via the Hubbard-Stratonovich
transformation \bqa && Z =
\int{Dc_{i\sigma}D\varphi_{i}Dm_{i}D\vec{\Omega}_{i}D\Delta_{ij}}e^{-\int_{0}^{\beta}d\tau
L} , \nn && L =
\sum_{i\sigma}c_{i\sigma}^{\dagger}(\partial_{\tau} -
\mu)c_{i\sigma} -
t\sum_{\ij\sigma}(c_{i\sigma}^{\dagger}c_{j\sigma} + H. c.) \nn &&
+ \sum_{i}\Bigl( \frac{1}{U}\varphi_{i}^{2} -
i\varphi_{i}\sum_{\sigma}{c}^{\dagger}_{i\sigma}{c}_{i\sigma}
\Bigr) \nn && + \sum_{i}\Bigl( \frac{1}{U}m_{i}^{2} -
m_{i}\sum_{\sigma\sigma'}{c}^{\dagger}_{i\sigma}({\vec
\Omega}_{i}\cdot{\vec \tau})_{\sigma\sigma'}{c}_{i\sigma'} \Bigr)
\nn && -
\sum_{\ij}\bigl[\Delta_{ij}(c_{i\uparrow}^{\dagger}c_{j\downarrow}^{\dagger}
- c_{i\downarrow}^{\dagger}c_{j\uparrow}^{\dagger}) + H. c.\bigr]
+ \frac{1}{J}\sum_{\ij}|\Delta_{ij}|^{2} , \eqa where
$\varphi_{i}$ and $m_{i}$ are local charge and spin potentials,
respectively.

The saddle-point analysis of Eq. (7) reveals its phase diagram
(Fig. 1) with a coexisting phase of antiferromagnetism and d-wave
superconductivity (AF$+$dSC), a d-wave superconducting state
(dSC), an itinerant antiferromagnetic phase (AF$+$FL), and a Fermi
liquid state (FL) in the plane of $(\delta, U/D)$ with hole
concentration $\delta$ and half bandwidth $D$. In this treatment
d-wave superconductivity competes with antiferromagnetism. In
addition, effects of Hubbard-U interactions can be incorporated
only via the antiferromagnetic order parameter. This means that
the pairing order parameter does not depend on the Hubbard-U when
the antiferromagnetic order parameter vanishes. Considering the
fact that the superconducting order parameter vanishes at the
point where four phases meet each other, it would also disappear
below this point owing to the independence of the pairing order
for U, and this separates the superconducting phase from the Fermi
liquid state via the straight line which is just a mean-field
artifact.
This argument seems to be inconsistent with the BCS model
analysis. When the antiferromagnetic order disappears, effects of
local interactions are not incorporated in this mean-field
treatment. Then, the effective Hamiltonian below the red-dotted
and red-dashed lines corresponds to the BCS model, which may
exhibit superconductivity for all fillings. Actually, however, the
pairing interaction strength will decrease effectively as hole
concentration  increases. The pairing order parameter with $m=0$
is shown in a renormalized mean-field theoretical
framework.\cite{RMT}

In this paper
our interest lies in the small doping region where
antiferromagnetic correlations would play an important role as
discussed in the slave-fermion context. We would like to emphasize
that our objective is to show the possible existence of an
intermediate paramagnetic metal with gapped spin excitations
between the collinear antiferromagnetic and d-wave superconducting
phases.
%
The BCS-HF analysis does not take into account
charge\cite{Kim_SU2_Rotor} and spin fluctuations as the heart of
Mott physics, particularly, in the low doping region. As a result,
the BCS-HF phase diagram does not allow such an intermediate state
between the antiferromagnetic Mott insulator and d-wave
superconductor. This is inconsistent with the slave-fermion
framework which is a strong coupling approach for the doped
antiferromagnetic Mott insulator.

\subsection{CP$^1$ representation with pairing fluctuations}

Our objective is to introduce spin fluctuations in the BCS-HF
effective theory [Eq. (7)]. A standard approach is to integrate
out electron degrees of freedom and obtain an effective action for
such spin fluctuations.\cite{HMM} In this context one can evaluate
the self-energy of electrons interacting with spin fluctuations.
Although this Landau-Ginzburg-Wilson-type approach is not
self-consistent, one can improve this methodology
to be self-consistent performing the Eliashberg-type
analysis.\cite{Spin_fluctuation_theory} Actually, this
spin-fluctuation theoretical framework has been applied to quantum
phase transitions of interacting itinerant electrons.\cite{FM_QCP}

However, such a spin-fluctuation theoretical framework is a
Fermi-liquid based weak-coupling approach, thus the resulting
normal state away from an antiferromagnetic phase is a Fermi
liquid state. This is in contrast with the slave-fermion framework
of the t-J model, allowing an anomalous metallic state far from
the Fermi liquid phase. One optimist may argue that the
spin-fluctuation approach can allow such a non-Fermi liquid phase
without symmetry breaking if interactions are taken into account
more heavily. Unfortunately, this kind of theoretical frameworks
have shown only symmetry breaking phases or the Fermi liquid state
if symmetric.\cite{Shankar_RG} In particular, the Fermi liquid
state was shown to be too stable, even up to two-loop calculations
in the renormalization group analysis,\cite{Shankar_RG} to evolve
other symmetric metallic phases. In this framework such an
anomalous metallic physics can appear only near its quantum
critical point. We do not claim that the spin-fluctuation approach
is not appropriate for studying the doped Mott insulator. We would
like to find the connection with the slave-fermion framework,
incorporating spin fluctuations into the BCS-HF effective theory.
Since our physical motivation lies in finding the anomalous
metallic state, claimed to arise in the slave-fermion
framework,\cite{Shankar,Slave_Fermion,Comment_SF} we develop a
strong coupling approach to allow spin fluctuations in the BCS-HF
theory.

To incorporate spin fluctuations into the BCS-HF theory [Eq. (7)]
in the context of the strong coupling approach, we resort to the
CP$^{1}$ spin-decomposition $\vec{\Omega}_{i}\cdot{\vec \tau} =
U_{i}\tau_{3}U^{\dagger}_{i}$, where $U_{i} = \left(
\begin{array}{cc} z_{i\uparrow} & - z_{i\downarrow}^{\dagger} \\
z_{i\downarrow} & z_{i\uparrow}^{\dagger} \end{array} \right)$ is
the SU(2) matrix field with the bosonic spinon
$z_{i\sigma}$.\cite{CP1,Kondo_Kim_Kim} Introducing the composite
field \bqa && \psi_{i\sigma} =
U^{\dagger}_{i\sigma\sigma'}c_{i\sigma'} \eqa in the strong
coupling approach,\cite{CP1,Kondo_Kim_Kim} Eq. (7) can be
expressed as \bqa && Z =
\int{D\psi_{i\sigma}Dz_{i\sigma}D\varphi_{i}Dm_{i}D\Delta_{ij}}
\delta(|z_{i\sigma}|^{2}-1) \nn &&
\exp\Bigl[-\int_{0}^{\beta}{d\tau} \Bigl\{
\frac{1}{U}\sum_{i}\varphi_{i}^{2} + \frac{1}{U}\sum_{i} m_{i}^{2}
+ \frac{1}{J}\sum_{\ij}|\Delta_{ij}|^{2} \nn && +
\sum_{i\sigma\sigma'}\psi_{i\sigma}^{\dagger}([\partial_{\tau} -
\mu - i\varphi_{i}]\delta_{\sigma\sigma'} +
[U^{\dagger}_{i}\partial_{\tau}U_{i}]_{\sigma\sigma'} \nn && -
m_{i} \tau^{3}_{\sigma\sigma'})\psi_{i\sigma'} -
t\sum_{\ij}\sum_{\sigma\sigma'\alpha}(\psi_{i\sigma}^{\dagger}U^{\dagger}_{i\sigma\alpha}U_{j\alpha\sigma'}\psi_{j\sigma'}
+ H.c.) \nn && -
\sum_{\ij}\Delta_{ij}[(z_{i\uparrow}^{\dagger}z_{j\downarrow}^{\dagger}-
z_{i\downarrow}^{\dagger}z_{j\uparrow}^{\dagger})\psi_{i\uparrow}^{\dagger}\psi_{j\uparrow}^{\dagger}
\nn&&-(z_{i\downarrow}z_{j\uparrow}-z_{i\uparrow}z_{j\downarrow})\psi_{i\downarrow}^{\dagger}\psi_{j\downarrow}^{\dagger}
+(z_{i\uparrow}^{\dagger}z_{j\uparrow}+z_{i\downarrow}^{\dagger}z_{j\downarrow})\psi_{i\uparrow}^{\dagger}\psi_{j\downarrow}^{\dagger}
\nn&&-(z_{i\downarrow}z_{j\downarrow}^{\dagger}+z_{i\uparrow}z_{j\uparrow}^{\dagger})\psi_{i\downarrow}^{\dagger}\psi_{j\uparrow}^{\dagger}
] - H.c. \Bigr\}\Bigr] . \eqa In this strong coupling
representation an antiferromagnetic spin fluctuation
$\vec{\Omega}_{i}$ carrying spin quantum number $1$ fractionalizes
into bosonic spinons $z_{i\sigma}$ with spin $1/2$, which seems to
occur through the screening of mobile electrons in the
antiferromagnetically correlated spin background. The components
of $\psi_{i\sigma}$ field are given by $\psi_{i\sigma} =
\left(\begin{array}{c} \psi_{i\uparrow}
\\ \psi_{i\downarrow} \end{array} \right) = \left(\begin{array}{c}
z^{\dagger}_{i\uparrow}c_{i\uparrow}
+  z_{i\downarrow}^{\dagger}c_{i\downarrow} \\
-z_{i\downarrow}c_{i\uparrow} + z_{i\uparrow}c_{i\downarrow}
\end{array} \right)$, which means that mobile electrons in the
antiferromagnetically correlated spin background fractionalize
into bosonic spinons $U_{i\sigma\sigma'}$ and fermionic chargons
$\psi_{i\sigma}$, i.e., $c_{i\sigma} =
U_{i\sigma\sigma'}\psi_{i\sigma'}$ in the strong coupling context.
An important observation in this representation is that fermion
pairing excitations couple to bosonic spin fluctuations, implying
that superconductivity is strongly correlated with
antiferromagnetism. As discussed in the previous section, it is
difficult for the slave-fermion approach to describe such pairing
interactions between charge degrees of freedom.

To make the present decomposition scheme natural, it is necessary
to understand the present methodology more deeply by comparing
this with other well studied ones. A good example is the quantum
disordered d-wave superconductivity for high $T_c$
cuprates,\cite{CP1_QDSC} where the coupling term of
$|\Delta|e^{i\phi}c_{\uparrow}c_{\downarrow}$ between Cooper pairs
and electrons plays the same role as the exchange coupling term of
${\vec \Omega}\cdot{c}_{\sigma}^{\dagger}{\vec
\tau}_{\sigma\sigma'}c_{\sigma'}$ between spin fluctuations and
electrons. Here, $|\Delta|$ and $\phi$ are the amplitude and phase
of Cooper pair fields. To solve this coupling term, several kinds
of gauge transformations are introduced. In these decoupling
schemes strong phase fluctuations of Cooper pairs, arising from
the phase-density uncertainty, screen out charge degrees of
freedom of electrons, causing electrically neutral but spinful
electrons called "spinons". As a result, the phase factor
disappears in the coupling term when it is rewritten in terms of
spinons. Instead, this coupling effect appears as current-current
interactions of neutral spinons and phase fields of Cooper pairs
in the kinetic term of electrons. Depending on the gauge
transformations, either Z$_2$ or U(1) gauge fields are obtained.
In this respect the present gauge transformation naturally extends
the methodology of charge U(1) symmetry in the context of
superconductivity to that of spin SU(2) symmetry in the context of
antiferromagnetism.

The "correlated" hopping term can be decomposed in the following
way \bqa && \exp\Bigl[- \int_{0}^{\beta}{d\tau} \Bigl\{
\sum_{i\sigma\sigma'}\psi_{i\sigma}^{\dagger}(\partial_{\tau}\delta_{\sigma\sigma'}
+ [U^{\dagger}_{i}\partial_{\tau}U_{i}]_{\sigma\sigma'}
)\psi_{i\sigma'} \nn && - t\sum_{\ij}\sum_{\sigma\sigma'\alpha}
(\psi_{i\sigma}^{\dagger}U^{\dagger}_{i\sigma\alpha}U_{j\alpha\sigma'}\psi_{j\sigma'}
+ H.c.) \Bigr\} \Bigr] \nn && \approx \exp\Bigl[- \Bigl\{ -
\sum_{i\tau\tau'}\sum_{\sigma\sigma'\alpha}
\psi_{i\tau}^{\dagger\sigma}U^{\dagger\sigma\alpha}_{i\tau}U_{i\tau'}^{\alpha\sigma'}\psi_{i\tau'}^{\sigma'}
+
\sum_{\tau}\sum_{i\sigma}\psi_{i\tau}^{\dagger\sigma}\psi_{i\tau}^{\sigma}
\nn&&-\frac{t}{J_{\tau}}\sum_{\ij\tau}\sum_{\sigma\sigma'\alpha}(
\psi_{i\tau}^{\dagger\sigma}U^{\dagger\sigma\alpha}_{i\tau}U_{j\tau}^{\alpha\sigma'}\psi_{j\tau}^{\sigma'}
+ H.c.) \Bigr\}\Bigr] \nn && =
\int{DF_{\mu\nu}^{\sigma\sigma'}DE_{\mu\nu}^{\sigma\sigma'}}\exp\Bigl[-
\sum_{i\tau\tau'}\Bigl\{
E_{i\tau\tau'}^{\dagger{\sigma\sigma'}}F_{i\tau\tau'}^{\sigma'\sigma}
+ H.c. \nn && -
U_{i\tau}^{{\dagger}\sigma\alpha}U_{i\tau'}^{\alpha\sigma'}F_{i\tau\tau'}^{\sigma'\sigma}
-
E_{i\tau\tau'}^{\dagger\sigma\sigma'}\psi_{i\tau'}^{\sigma'}\psi_{i\tau}^{\dagger\sigma}
- H.c. \Bigr\}\nn && -
\sum_{\tau}\sum_{i\sigma}\psi_{i\tau}^{\dagger\sigma}\psi_{i\tau}^{\sigma}
- \frac{t}{J_{\tau}}\sum_{\ij\tau}\Bigl\{
E_{ij\tau}^{\dagger{\sigma\sigma'}}F_{ij\tau}^{\sigma'\sigma} +
H.c. \nn && -
U_{i\tau}^{\dagger\sigma\alpha}U_{j\tau}^{\alpha\sigma'}F_{ij\tau}^{\sigma'\sigma}
-
E_{ij\tau}^{\dagger\sigma\sigma'}\psi_{j\tau}^{\sigma'}\psi_{i\tau}^{\sigma\dagger}
- H.c. \Bigr\}\Bigr] , \eqa where the time part in the effective
hopping term is evaluated in the discrete-time approximation with
$J_{\tau}$, an energy scale for discrete time. This kind of
approximation has been well adopted in the Monte Carlo simulation,
known as the Suzuki-Trotter decomposition.\cite{MC_ST} Its
detailed derivation can be found in Ref. \cite{Kondo_Kim_Kim}.

The effective hopping parameters can be represented as their
amplitudes and phases \bqa &&
E_{i\tau\tau'}^{\dagger{\sigma\sigma'}} \equiv
E_{\tau}e^{-ic_{i\tau\tau'}\tau_{3\sigma\sigma'}} , ~~~~~
F_{i\tau\tau'}^{\sigma'\sigma}  \equiv
F_{\tau}e^{ic_{i\tau\tau'}\tau_{3\sigma\sigma'}} , \nn &&
E_{ij\tau}^{\dagger{\sigma\sigma'}} \equiv
E_{r}e^{-ic_{ij\tau}\tau_{3\sigma\sigma'}} , ~~~~~
F_{ij\tau}^{\sigma'\sigma}  \equiv
F_{r}e^{ic_{ij\tau}\tau_{3\sigma\sigma'}} , \eqa where the unknown
amplitudes are determined self-consistently in the saddle-point
analysis as $E_{\tau} = |\langle
U_{i\tau}^{{\dagger}\sigma\alpha}U_{i\tau'}^{\alpha\sigma'}
\rangle|$, $F_{\tau} = |\langle
\psi_{i\tau'}^{\sigma'}\psi_{i\tau}^{\dagger\sigma} \rangle|$,
$E_{r} = |\langle
U_{i\tau}^{\dagger\sigma\alpha}U_{j\tau}^{\alpha\sigma'} \rangle|$
, and $F_{r} = |\langle
\psi_{j\tau}^{\sigma'}\psi_{i\tau}^{\sigma\dagger} \rangle|$.

The correlated pairing term can be expressed as \bqa && -
\sum_{\ij}\Delta_{ij}[(z_{i\uparrow}^{\dagger}z_{j\downarrow}^{\dagger}-
z_{i\downarrow}^{\dagger}z_{j\uparrow}^{\dagger})\psi_{i\uparrow}^{\dagger}\psi_{j\uparrow}^{\dagger}
-(z_{i\downarrow}z_{j\uparrow}-z_{i\uparrow}z_{j\downarrow})\psi_{i\downarrow}^{\dagger}\psi_{j\downarrow}^{\dagger}
\nn&&+(z_{i\uparrow}^{\dagger}z_{j\uparrow}+z_{i\downarrow}^{\dagger}z_{j\downarrow})\psi_{i\uparrow}^{\dagger}\psi_{j\downarrow}^{\dagger}
-(z_{i\downarrow}z_{j\downarrow}^{\dagger}+z_{i\uparrow}z_{j\uparrow}^{\dagger})\psi_{i\downarrow}^{\dagger}\psi_{j\uparrow}^{\dagger}
] - H.c. \nn && \approx - \sum_{\ij}\Delta_{ij}\Bigl[E_{r}
(e^{-ic_{ij}}\psi_{i\uparrow}^{\dagger}\psi_{j\downarrow}^{\dagger}
-
e^{ic_{ij}}\psi_{i\downarrow}^{\dagger}\psi_{j\uparrow}^{\dagger})
\nn&&- E_{r}\Delta_{ij}^{\psi}\cos(\vartheta_{ij}^{\psi} - c_{ij})
+ \frac{\Delta_{ij}^{\psi}}{2}\sum_{\sigma}
(z_{i\sigma}^{\dagger}e^{i\vartheta_{ij}^{\psi}}z_{j\sigma} + H.
c. ) \Bigr] \nn && - H.c. \eqa in the "saddle-point"
approximation, where the mean-field ansatz of
$\langle\psi_{i\uparrow}^{\dagger}\psi_{j\downarrow}^{\dagger}\rangle
=  \frac{\Delta_{ij}^{\psi}}{2}e^{i\vartheta^{\psi}_{ij}}$,
$\langle\psi_{i\downarrow}^{\dagger}\psi_{j\uparrow}^{\dagger}\rangle
= - \frac{\Delta_{ij}^{\psi}}{2}e^{-{i\vartheta^{\psi}_{ij}}}$,
$\langle\psi_{i\uparrow}^{\dagger}\psi_{j\uparrow}^{\dagger}\rangle
= 0$, and
$\langle\psi_{i\downarrow}^{\dagger}\psi_{j\downarrow}^{\dagger}\rangle
= 0$ is utilized. The different signs of the phase factor
originate from the U(1) gauge symmetry associated with the
CP$^{1}$ representation\cite{Gauge_Transformation} while that of
the pairing amplitude recovers singlet pairing. It is important to
notice that the mean-field ansatz for the pairing sector is
consistent with that for the hopping sector.

Performing the continuum approximation for the time
part,\cite{Kondo_Kim_Kim} we obtain the following expression for
the effective Lagrangian \bqa && L_{eff} = \frac{1}{U}\sum_{i}
(\varphi_{i}^{2} + m_{i}^{2}) +
\frac{1}{J}\sum_{\ij}|\Delta_{ij}|^{2} + \mu\sum_{i}(1-\delta) \nn
&& +
\sum_{\ij}E_{r}\Delta_{ij}\Delta_{ij}^{\psi}\cos(\vartheta_{ij}^{\psi}
- c_{ij}) + J_{\tau}\sum_{i}(E_{\tau}-1)F_{\tau} \nn && +
t\sum_{\ij}E_{r}F_{r} +
\sum_{i\sigma}\psi_{i\sigma}^{\dagger}(E_{\tau}[\partial_{\tau} -
i\sigma{c}_{i\tau}] - i\varphi_{i} - \sigma{m}_{i})\psi_{i\sigma}
\nn && - (\mu +
J_{\tau}[E_{\tau}-1])\sum_{i\sigma}\psi_{i\sigma}^{\dagger}\psi_{i\sigma}
\nn && -
t{E}_{r}\sum_{\ij\sigma}(\psi_{i\sigma}^{\dagger}e^{-i\sigma{c}_{ij}}\psi_{j\sigma}
+ H.c.) \nn &&  - \sum_{\ij}E_{r}\Delta_{ij}
(e^{-ic_{ij}}\psi_{i\uparrow}^{\dagger}\psi_{j\downarrow}^{\dagger}
-
e^{ic_{ij}}\psi_{i\downarrow}^{\dagger}\psi_{j\uparrow}^{\dagger})
- H.c. \nn && +
\frac{F_{\tau}}{J_{\tau}}\sum_{i\sigma}|(\partial_{\tau} -
ic_{i\tau})z_{i\sigma}|^{2} - t {F}_{r}\sum_{\ij\sigma}
(z_{i\sigma}^{\dagger}e^{ic_{ij}}z_{j\sigma} + H.c.) \nn && -
\sum_{\ij\sigma}\Delta_{ij}\Delta_{ij}^{\psi}z_{i\sigma}^{\dagger}e^{i\vartheta_{ij}^{\psi}}z_{j\sigma}
- H.c. + i\sum_{i}\lambda_{i}(\sum_{\sigma}|z_{i\sigma}|^{2} - 1)
, \nn \eqa where $\lambda_{i}$ is a Lagrange multiplier field
imposing the uni-modular constraint. It is important to observe
the new energy scale $J_{\tau}/F_{\tau}$ in the boson sector for
spin dynamics. It is well known that the energy scale for spin
dynamics is different from the Hubbard-U. In this paper we simply
assume $F_{\tau}/J_{\tau} = 1/g$ for spin dynamics and $E_{\tau} =
1$ for charge dynamics, where $g$ is an effective coupling
constant for spin fluctuations, although the full analysis with
$F_{\tau}$ and $E_{\tau}$ is possible. This simplification would
not alter the phase structure of the present effective theory. It
is expected that the
$E_{r}\Delta_{ij}\Delta_{ij}^{\psi}\cos(\vartheta_{ij}^{\psi} -
c_{ij})$ term in the pairing sector [Eq. (12)] is relevant at low
energies, allowing us to set $\vartheta_{ij}^{\psi} = c_{ij} +
\pi$. Replacing $t {F}_{r} - \Delta_{ij}\Delta_{ij}^{\psi}$ with
$tF_{r}$, the pseudo-fermion pairing order parameter
$\Delta_{ij}^{\psi}$ disappears in the effective Lagrangian, and
only the electron pairing order parameter $\Delta_{ij}$ appears,
consistent with our expectation.

We find the effective Lagrangian for the doped antiferromagnetic
Mott insulator \bqa && L_{eff} = L_{0} + L_{BCS-HF} +
L_{NL\sigma{M}} , \nn && L_{0} = \frac{1}{U}\sum_{i}
(\varphi_{i}^{2} + m_{i}^{2}) +
\frac{1}{J}\sum_{\ij}|\Delta_{ij}|^{2} + t\sum_{\ij}E_{r}F_{r} \nn
&& + \mu\sum_{i}(1-\delta) , \nn && L_{BCS-HF} =
\sum_{i\sigma}\psi_{i\sigma}^{\dagger}( \partial_{\tau} -
i\sigma{c}_{i\tau} - \mu - i\varphi_{i} -
\sigma{m}_{i})\psi_{i\sigma} \nn && -
t{E}_{r}\sum_{\ij\sigma}(\psi_{i\sigma}^{\dagger}e^{-i\sigma{c}_{ij}}\psi_{j\sigma}
+ H.c.) \nn && - \sum_{\ij}E_{r}\Delta_{ij}
(e^{-ic_{ij}}\psi_{i\uparrow}^{\dagger}\psi_{j\downarrow}^{\dagger}
-
e^{ic_{ij}}\psi_{i\downarrow}^{\dagger}\psi_{j\uparrow}^{\dagger})
- H.c. , \nn && L_{NL\sigma{M}} =
\frac{1}{g}\sum_{i\sigma}|(\partial_{\tau} -
ic_{i\tau})z_{i\sigma}|^{2} \nn && - tF_{r} \sum_{\ij\sigma}
(z_{i\sigma}^{\dagger}e^{ic_{ij}}z_{j\sigma} + H.c.) +
i\sum_{i}\lambda_{i}(\sum_{\sigma}|z_{i\sigma}|^{2} - 1) . \nn
\label{SRGT} \eqa Spin dynamics of the doped antiferromagnetic
Mott insulator is governed by the CP$^1$ gauge theory of the
nonlinear $\sigma$ model with the renormalized spinon-bandwidth
$DF_{r}$. On the other hand, the fermion sector describing charge
dynamics coincides with the BCS-HF theory [Eq. (7)] except the
renormalized pairing order parameter $E_{r}\Delta_{ij}$ with the
renormalized bandwidth $DE_{r}$, ignoring spin-gauge fluctuations
$c_{ij}$ in the saddle-point approximation. As a result, we can
describe the spin and charge dynamics of the doped
antiferromagnetic Mott insulator with three order parameters given
by $\langle{z}_{i\sigma}\rangle$, $m_{i}$, and $\Delta_{ij}$.

Since fermion excitations are gapped at half filling, spin
fluctuations are only relevant degrees of freedom at low energies,
and their dynamics is governed by the O(3) nonlinear $\sigma$
model. Thus, one theoretical limit of the half-filled case is
recovered correctly. Away from half filling, charge fluctuations
can be gapless to show metallic properties because the chemical
potential shifts from the middle of the BCS-HF gap to the lower
band exhibiting finite density of states. In addition, one can
find that spin-boson excitations are condensed at moderate values
of U. These condensed spinons are confined with fermions to form
coherent electron excitations, consistent with the conventional BCS-HF
theory. Recall that spin-gauge fluctuations are gapped due to the
Anderson-Higgs mechanism. Further hole doping would lead the
antiferromagnetic order to vanish while the pairing order
survives. The resulting superconducting phase is described by the
BCS theory, consistent with the other theoretical limit away from
half filling.

An interesting question is whether spinon excitations can be
gapped to cause spin-gauge fluctuations gapless in the
intermediate doping region between the antiferromagnetic Mott
insulator and d-wave superconductor. Since spin-gauge fluctuations
play the role of phase fluctuations of renormalized Cooper pairs
($E_{r}\Delta_{ij}$), this intermediate phase would be a
spin-gapped metal with preformed pairing. Such an anomalous metal
is the primary discovery of the present paper.

\subsection{Connection with the slave-fermion approach}

It is valuable to find the connection between the slave-fermion
approach and present theoretical framework. Although its precise
connection is not easy to construct, one can understand the
relationship qualitatively. It was shown that the slave-fermion
Lagrangian at half filling, more precisely the Schwinger-boson
theory can recover the CP$^1$ Lagrangian of the nonlinear $\sigma$
model in the long-wave length and low energy
limits,\cite{Sachdev_NLsM} where the hopping parameter
$\chi_{ij}^{b}$ vanishes and the fermion dynamics disappears in
the slave-fermion Lagrangian of Eq. (4), that is, $L_{eff} =
\sum_{i\sigma}b_{i\sigma}^{\dagger}(\partial_{\tau} + i
\lambda_{i})b_{i\sigma} -
\sum_{\ij\sigma\sigma'}(\Delta_{ij}^{\dagger}\epsilon_{\sigma\sigma'}
b_{i\sigma}b_{j\sigma'} + H.c.)$ without constant terms.
Diagonalizing the effective Hamiltonian, one finds the dispersion
relation of boson excitations. Considering such boson excitations
around the energy minima and phase fluctuations of singlet-pairing
excitations, one can find the CP$^1$ gauge theory from the above
Schwinger-boson Lagrangian in the low energy limit. Here, the
$z_{i\sigma}$ field consists of particle-hole linear-combination
of the Schwinger-boson field, and the CP$^1$ gauge field arises
from the phase field of the singlet-pairing order
parameter.\cite{Sachdev_NLsM}

Hole doping will give rise to a nonzero hopping parameter
$\chi_{ij}^{b}$. One can solve such an effective Hamiltonian in
the same strategy as the Schwinger-boson case. Actually, one of
the present authors is preforming the self-consistent analysis in
the presence of fermion excitations.\cite{Comment_SF,Kim_Jia} One
clear point is that the boson dynamics is relativistic, thus the
Klein-Gordon-type Lagrangian will be obtained for low energy
dynamics of such boson excitations, corresponding to the CP$^1$
gauge Lagrangian if irrelevant terms are abandoned appropriately.

For fermion dynamics, it is more difficult to find its connection
since there are no particle-particle and particle-hole pairing
fluctuations in the slave-fermion representation of the t-J model.
Even if pairing excitations are neglected in Eq. (14), it is
difficult to make the "spin"-dependent chemical potential, $\sigma
m$ of Eq. (14) in the slave-fermion Lagrangian of Eq. (4). In
addition, there is no spin index in $L_{SF}$ of Eq. (4). One
possible way to overcome this inconsistency is to take the low
energy limit of Eq. (14). If the long-wave length and low energy
limits are considered in Eq. (14), only one flavor of fermions
will appear. Then, the resulting low-energy fermion-gauge
Lagrangian with a Fermi surface is expected to be basically the
same as that of the slave-fermion representation. In section V we
discuss possible low-energy effective Lagrangians in both cases,
and argue that the spin-gapped metallic phase where boson
excitations are gapped is identified with the same fixed point of
the same effective fermion-gauge Lagrangian.

The above discussion tells us that the CP$^1$ decomposition
approach of the HF effective model without pairing fluctuations
will share the similar physics with the slave-fermion framework of
the t-J model, in particular, when bosonic spin fluctuations are
gapped. However, there is one difficulty in the present approach.
The present decomposition scheme does not have any small
parameters, in contrast with the slave-fermion framework where the
spin index can be extended as $\sigma = 1 , ..., N$, thus allowing
the $1/N$ expansion. Since the CP$^1$ decomposition can be allowed
only in the case of $N = 2$, it is not easy to justify its
saddle-point analysis against gauge fluctuations. This is the
reason why we compare the present framework with the slave-fermion
approach, where the mean-field analysis can be justified in the
$1/N$ expansion. There is one more possibility to make the present
saddle-point analysis stable against gauge fluctuations. Since
gauge fluctuations are dissipative due to the presence of the
Fermi surface, strong damping in gauge fluctuations may give rise
to the stability of the mean-field
analysis.\cite{Deconfinement_Nonrelativistic1} Actually, one of
the present authors has discussed that average gauge fluctuations
are proportional to $1/\sigma_{f}$, implying that such
fluctuations will be suppressed in the infinite limit of the
fermion conductivity $\sigma_{f}$ and allowing the mean-field
analysis stable against gauge
fluctuations.\cite{Deconfinement_Nonrelativistic1} This important
issue is intensively discussed in section V.

\section{Mean-field analysis and phase diagram}

\subsection{Phase diagram}

Taking the mean-field ansatz of antiferromagnetism [$m_{i} =
(-1)^{i}m$] and d-wave pairing [$\Delta_{ij} = \Delta$ with $j = i
\pm {\hat{x}}$ and $\Delta_{ij} = - \Delta$ with $j = i \pm
{\hat{y}}$] with $i\varphi_{i} = \varphi$ and $i \lambda_{i} =
\lambda$, we obtain the free energy functional from Eq. (14) \bqa
&& F_{MF} = \sum_{k}\Bigl( \frac{- \varphi^{2} + m^{2}}{U} +
\frac{\Delta^{2}}{2J} + DE_{r}F_{r} - \mu \delta - \varphi -
\lambda \Bigr) \nn && - \frac{1}{\beta}\sum_{\omega_{n}}\sum'_{k}
\sum_{s,s'=\pm}\ln(i\omega_{n} - E_{kss'}) \nn && +
\frac{1}{\beta}\sum_{\nu_{n}}\sum_{k\sigma}\ln\Bigl(\frac{1}{g}\nu_{n}^{2}
+ {F}_{r}\epsilon_{k} + \lambda \Bigr) . \eqa Here the
renormalized fermion spectrum is given by $E_{k\pm\pm} = \pm
\sqrt{\Bigl(\sqrt{(E_{r}\epsilon_{k})^{2} + m^{2}} \pm |\mu +
\varphi|\Bigr)^{2} + (E_{r}\Delta_{k})^{2}}$ with $\Delta_{k} =
\Delta(\cos k_{x} - \cos k_{y})$ and $\epsilon_{k} = - 2t(\cos
k_{x} + \cos k_{y})$. $\omega_{n}$ ($\nu_{n}$) is the Matzubara
frequency for fermions (bosons) with temperature $T \equiv
1/\beta$.

Minimizing the free energy [Eq. (15)] with respect to $m$,
$\Delta$, $E_{r}$, $F_{r}$, $\lambda$, $\varphi$, and $\mu$, we
obtain the self-consistent mean-field equations. Performing the
Matzubara frequency summations and momentum integrals with
$\sum_{k} = \int_{-D}^{D}d\epsilon D(\epsilon)$ and $D(\epsilon) =
1/ (2D)$ for the boson sector, we find \bqa && DF_{r} = \sum_{k}'
\Bigl[ \frac{\Bigl(\sqrt{(E_{r}\epsilon_{k})^{2} + m^{2}} +
\mu_{r}\Bigr) \frac{E_{r}\epsilon_{k}^{2}}{
\sqrt{(E_{r}\epsilon_{k})^{2} + m^{2}}} + E_{r}\Delta_{k}^{2}
}{\sqrt{\Bigl(\sqrt{(E_{r}\epsilon_{k})^{2} + m^{2}} +
\mu_{r}\Bigr)^{2} + (E_{r}\Delta_{k})^{2}} } \nn && +
\frac{\Bigl(\sqrt{(E_{r}\epsilon_{k})^{2} + m^{2}} - \mu_{r}\Bigr)
\frac{E_{r}\epsilon_{k}^{2}}{ \sqrt{(E_{r}\epsilon_{k})^{2} +
m^{2}}} + E_{r}\Delta_{k}^{2}
}{\sqrt{\Bigl(\sqrt{(E_{r}\epsilon_{k})^{2} + m^{2}} -
\mu_{r}\Bigr)^{2} + (E_{r}\Delta_{k})^{2}} } \Bigr] , \nn &&
\frac{2m}{U}  = \sum_{k}' \frac{m}{ \sqrt{(E_{r}\epsilon_{k})^{2}
+ m^{2}}} \nn && \Bigl[ \frac{\Bigl(\sqrt{(E_{r}\epsilon_{k})^{2}
+ m^{2}} +
\mu_{r}\Bigr)}{\sqrt{\Bigl(\sqrt{(E_{r}\epsilon_{k})^{2} + m^{2}}
+ \mu_{r}\Bigr)^{2} + (E_{r}\Delta_{k})^{2}} } \nn && +
\frac{\Bigl(\sqrt{(E_{r}\epsilon_{k})^{2} + m^{2}} -
\mu_{r}\Bigr)}{\sqrt{\Bigl(\sqrt{(E_{r}\epsilon_{k})^{2} + m^{2}}
- \mu_{r}\Bigr)^{2} + (E_{r}\Delta_{k})^{2}} } \Bigr]  , \nn &&
\frac{\Delta}{J}  = \sum_{k}' \Delta \Bigl[ \frac{E_{r}^{2} (\cos
k_x - \cos k_y)^{2}}{\sqrt{\Bigl(\sqrt{(E_{r}\epsilon_{k})^{2} +
m^{2}} + \mu_{r}\Bigr)^{2} + (E_{r}\Delta_{k})^{2}} }  \nn && +
\frac{E_{r}^{2} (\cos k_x - \cos
k_y)^{2}}{\sqrt{\Bigl(\sqrt{(E_{r}\epsilon_{k})^{2} + m^{2}} -
\mu_{r}\Bigr)^{2} + (E_{r}\Delta_{k})^{2}} } \Bigr] , \nn &&
\delta  = \sum_{k}' \Bigl[
\frac{\Bigl(\sqrt{(E_{r}\epsilon_{k})^{2} + m^{2}} +
\mu_{r}\Bigr)}{\sqrt{\Bigl(\sqrt{(E_{r}\epsilon_{k})^{2} + m^{2}}
+ \mu_{r}\Bigr)^{2} + (E_{r}\Delta_{k})^{2}} }  \nn && -
\frac{\Bigl(\sqrt{(E_{r}\epsilon_{k})^{2} + m^{2}} -
\mu_{r}\Bigr)}{\sqrt{\Bigl(\sqrt{(E_{r}\epsilon_{k})^{2} + m^{2}}
- \mu_{r}\Bigr)^{2} + (E_{r}\Delta_{k})^{2}} } \Bigr] \eqa for the
fermion part with $\varphi = - \frac{U}{2}(1-\delta)$ and $\mu_{r}
= \mu + \varphi$, and \bqa && 1 =
\frac{\sqrt{\lambda+DF_{r}}-\sqrt{\lambda-DF_{r}}}{\sqrt{1/g}(DF_{r})}
, \nn && E_{r} =
\frac{(2\lambda-DF_{r})\sqrt{\lambda+DF_{r}}-(2\lambda+DF_{r})\sqrt{\lambda-DF_{r}}}{3\sqrt{1/g}(DF_{r})^{2}}
\nn \eqa for the spinon sector.

\begin{figure}[t]
\vspace*{7cm} \includegraphics{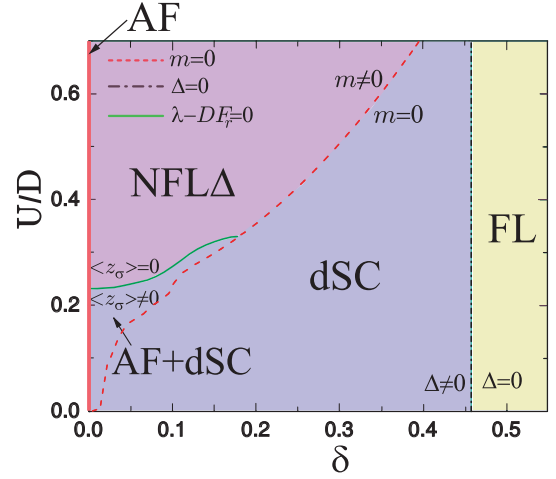} \vspace{-0.5cm}
\caption{(Color online) Mean
field phase diagram of the spin decomposition theory of Eq. (14)
with $J/D=0.7$ and $g=0.42$ at zero temperature. Separation
between the dSC and FL via the straight line is an artifact of the
BCS-HF analysis for the fermion sector.
Introduction of spin fluctuations in the
BCS-HF theory alters the AF$+$dSC phase of the BCS-HF phase
diagram into the spin-gapped non-Fermi liquid state with d-wave
pairing fluctuations (NFL$\Delta$) since such spin fluctuations
are gapped in the strong coupling analysis. See the text.}
\label{Fig2}
\end{figure}

The resulting phase diagram is shown in Fig. 2. The HF
phase-boundary characterized by $m = 0$ is qualitatively similar
with that of Eq. (7) [Fig. 1] although the region of $m \not= 0$
in Fig. 2 is larger than that in the BCS-HF phase diagram, arising
from band renormalization $E_{r}D$ to increase the fermion density
of states. On the contrary to the BCS-HF phase diagram, we find
the region where d-wave pairing order does not exist for large
$U/D$ and low $\delta$. Such a region is not shown in order to
clarify the difference between Fig. 2 and Fig. 1, that is, the
emergence of a spin-gapped incoherent metal with preformed pairing
excitations denoted by NFL$\Delta$, as will be discussed below in
more detail. The absence of d-wave pairing in large interaction
and small doping originates from the band ($E_{r}D$) and pairing
($E_{r}\Delta$) renormalization due to spin fluctuations.
Actually, the spin-gapped anomalous metal without pairing
fluctuations can be found from the HF Lagrangian without the
pairing term, using the same strong coupling approach as the
CP$^1$ decomposition.

In Fig. 3 we show doping dependence of BCS-HF order parameters for
various $U/D$. The magnetization amplitude scaled by the half
bandwidth decreases from its maximum value at half filling as hole
concentration increases, exhibiting the second order transition.
The d-wave pairing order parameter shows an arch-like shape in the
parameter range of $U/D$, where it vanishes at half filling due to
competition with antiferromagnetism. The black dotted line denotes
the point where the magnetization amplitude vanishes, implying
that the pairing order parameter does not depend on the Hubbard
interaction $U/D$ from this hole concentration, as discussed in
the BCS-HF phase diagram [Fig. 1]. By the same reason as the
BCS-HF phase diagram the d-wave superconducting phase separates
from the Fermi liquid state via the straight line, but this is an
artifact of the BCS-HF mean-field analysis for the fermion sector.
This doping region is out of interest in the present paper, which we
will not focus on.

\begin{figure}[t]
\vspace*{9.5cm} \includegraphics{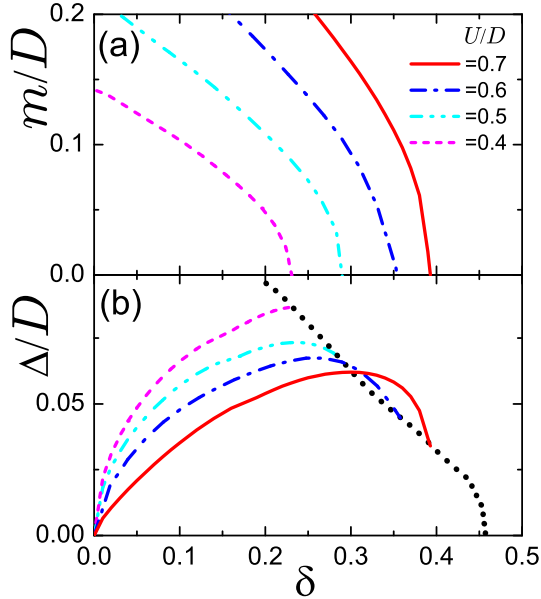} \vspace{-0.5cm}
\caption{(Color online) (a)
Magnetization amplitude and (b) d-wave pairing gap in the
NFL$\Delta$ phase of Fig. \ref{Fig2}. The black dotted line shows
the pairing gap calculated with $m=0$ in the dSC region. }
\label{Fig3}
\end{figure}

The main point in our spin-decomposition theory is that there is
an additional transition line associated with the condensation of
spinons. The condensation transition occurs when the boson
excitation gap closes, given by $\lambda_{c} - DF_{rc} = 0$ in Eq.
(17) where the subscript $c$ denotes the quantum critical point.
We find the condensation-transition point $(D/g) F_{rc} = 2$ with
$E_{rc} = 1/3$ from Eq. (17). Below this transition line the
spin-decomposition theory [Eq. (14)] is reduced to the BCS-HF
theory [Eq. (7)] owing to the spinon condensation (Higgs phase).
Thus, as $\delta$ increases below this line, the phase diagram
shows a coexistence region of antiferromagnetism and d-wave
superconductivity (AF$+$dSC: $m \not= 0$, $\Delta \not= 0$,
$\langle{z_{\sigma}}\rangle \not= 0$) and d-wave superconducting
state (dSC: $m = 0$, $\Delta \not= 0$, $\langle{z_{\sigma}}\rangle
\not= 0$), perfectly consistent with the BCS-HF phase diagram.
Here, the boundary line between AF$+$dSC and dSC is obtained by
extrapolation of $m$ in the NFL$\Delta$ region.\cite{AFdSC_dSC}
Above the transition line, where spin fluctuations are gapped,
symmetry-breaking patterns discriminate two non-Fermi liquid (NFL)
phases as $U/D$ increases: NFL$\Delta$ with pairing fluctuations
($m \not= 0$, $\Delta \not= 0$, $\langle{z_{\sigma}}\rangle = 0$)
and NFL ($m \not= 0$, $\Delta = 0$, $\langle{z_{\sigma}}\rangle =
0$), not shown in Fig. 2.

One cautious person may ask why the AF state is limited to appear
at half filling in the case of large $U/D$ although the mean-field
phase diagram [Fig. 2] shows such an antiferromagnetic state away
from half filling, coexisting with d-wave superconductivity in the
case of small $U/D$. This should be regarded as an artifact of the
mean-field analysis in the gauge theoretic description. We note
that the antiferromagnetic phase is characterized by not only the
magnetization amplitude $m$ but also its directional fluctuation
$\vec{\Omega}_{i}$. Since the magnetization amplitude is
determined by the conventional HF calculation, its nonzero region
covers large hole concentration, consistent with the BCS-HF phase
diagram [Fig. 1]. However, nonzero $m$ itself does not mean the
presence of the antiferromagnetic order, since directional
spin-fluctuations may break the magnetic order. When interactions
are weak, such directional fluctuations become suppressed. Then,
the antiferromagnetic order appears to coexist with the
superconducting order, consistent with the BCS-HF result.

As interactions increase, spin fluctuations become strong. Such
directional fluctuations are represented as fractionalized boson
excitations $z_{\sigma}$ in the strong coupling analysis (CP$^{1}$
decomposition). Even at half filling, such boson excitations can
be gapped in the strong coupling case. This is certainly an
artifact of the gauge theory approach, which usually occurs in the
mean-field calculation. If gauge fluctuations (instanton effects)
are taken into account appropriately, confinement should arise at
half filling.\cite{Polyakov,Fradkin,NaLee} Thus, the
antiferromagnetic order will be recovered at half filling as an
instanton effect (confinement) when the interaction is large. This
also happens in the pure nonlinear $\sigma$ model as the coupling
$g$ increases.\cite{Tsvelik,Nagaosa_book} Such gapped boson
excitations should be also confined via gauge interactions.

On the other hand, away from half filling, there emerge gapless
fermion excitations. The presence of gapless excitations can give
rise to deconfinement,\cite{Deconfinement_Nonrelativistic1,
Deconfinement_Relativistic_Boson,
Deconfinement_Relativistic_Fermion,Deconfinement_Nonrelativistic2}
thus such a disordered phase ($\langle{z}_{\sigma}\rangle = 0$
$\rightarrow$ $\langle{\vec \Omega}\rangle = 0$) may be
stabilized. However, the existence of such a deconfinement phase
depends on how many flavors of gapless matters there
are.\cite{Deconfinement_Nonrelativistic1,Deconfinement_Relativistic_Boson,
Deconfinement_Relativistic_Fermion,Deconfinement_Nonrelativistic2}
This means that, if hole concentration is small, the density of
gapless fermions may not be enough to allow
deconfinement.\cite{Deconfinement_Nonrelativistic1,Deconfinement_Nonrelativistic2}
In this case confinement can arise, and such a paramagnetic
anomalous metal ($m = 0$ and $\langle{z}_{\sigma}\rangle = 0$)
becomes unstable in the small $\delta$ and large U region. As a
result, the antiferromagnetic order can persist up to small but
finite hole concentration in large U. This antiferromagnetism can
be considered as the extension of the antiferromagnetic order at
half filling in large U (arising from confinement in the gauge
theory context) to a small doping region. Such a confinement issue
will be discussed more deeply in section V.
The present mean-field analysis overestimates directional
spin-fluctuations since gauge fluctuations are ignored.
Introduction of gauge fluctuations has been shown to increase
antiferromagnetic correlations.\cite{AF_Correlation} To determine
the critical hole concentration where the antiferromagnetic order
vanishes is certainly beyond the scope of this paper because it is
associated with the confinement issue far beyond the mean-field
description.

It is important to understand how the superconducting phase is
characterized in the spin-decomposition theory. The BCS
superconducting order parameter is not sufficient to confirm the
existence of superconductivity. As shown in the effective
Lagrangian Eq. (14), the fermion pairing term consists of $-
\sum_{\langle ij\rangle}E_{r}\Delta_{ij}
(e^{-ic_{ij}}\psi_{i\uparrow}^{\dagger}\psi_{j\downarrow}^{\dagger}
-
e^{ic_{ij}}\psi_{i\downarrow}^{\dagger}\psi_{j\uparrow}^{\dagger})$.
The point is the presence of the phase-fluctuation term
$e^{-ic_{ij}}$ arising from spin fluctuations. Thus, for
superconductivity to be truly realized, not only nonzero $\Delta$
but also $\langle e^{-ic_{ij}}\rangle \not= 0$ is required, where
$\langle e^{-ic_{ij}}\rangle \not= 0$ can be achieved by boson
condensation. Notice that the latter is nothing but the Higgs
mechanism since $c_{ij}$ in the phase factor corresponds to the
CP$^{1}$ gauge field. Thus, when such boson excitations are gapped
as shown in the mean-field analysis, phase fluctuations of fermion
pairs are strong, and the superconducting order does not appear.
This is the reason why the non-Fermi liquid phase with preformed
pairing fluctuations arises in the strong coupling analysis.

The NFL phase is basically the same as the spin-gapped metal in
the slave-fermion context, where bosonic spin degrees of freedom
are gapped, but fermionic charge degrees of freedom are gapless.
They are interacting via gauge fluctuations, thus exhibiting
anomalous metallic physics. On the other hand, the NFL$\Delta$
phase extends the slave-fermion framework, incorporating pairing
fluctuations into the slave-fermion scheme. Thus, the NFL state
turns into the NFL$\Delta$ phase in the intermediate $U/D$ above
the transition line after superconducting correlations are taken
into account. Remember that introduction of superconducting
correlations is not trivial in the slave-fermion context since
pairing interactions between charge degrees of freedom are not
allowed in the naive mean-field scheme.

\subsection{Physical implication}

Both the NFL$\Delta$ and NFL metallic states do not allow coherent
electron excitations and spin fluctuations owing to deconfined
gapped spinon excitations. Consider the electron Green's function
and spin susceptibility \bqa
G_{\uparrow\uparrow}^{el}(ij,\tau\tau') && \approx
g^{z}_{\uparrow}(ij,\tau\tau')g^{\psi}_{\uparrow}(ij,\tau\tau')
\nn && +
g^{z}_{\downarrow}(ji,\tau'\tau)g^{\psi}_{\downarrow}(ij,\tau\tau')
, \nn \chi^{zz}(ij,\tau\tau') && \approx
(1-\delta)^{2}[g_{\uparrow}^{z}(ji,\tau'\tau)g_{\uparrow}^{z}(ij,\tau\tau')
\nn && +
g_{\downarrow}^{z}(ji,\tau'\tau)g_{\downarrow}^{z}(ij,\tau\tau')]
, \eqa where each propagator is expressed as
$g^{z}_{\uparrow(\downarrow)}(ij,\tau\tau') =
\langle{T}_{\tau}[z_{i\tau\uparrow(\downarrow)}z_{j\tau'\uparrow(\downarrow)}^{\dagger}]\rangle$
and $g^{\psi}_{\uparrow(\downarrow)}(ij,\tau\tau') =
\langle{T}_{\tau}[\psi_{i\tau\uparrow(\downarrow)}\psi_{j\tau'\uparrow(\downarrow)}^{\dagger}]\rangle$,
respectively. It is important to understand that both response
functions consist of convolution integrals. As a result, only
"particle-hole" continuum spectrum can be observed when spin
excitations are fractionalized or deconfined.

Increasing $\delta$ with a fixed $U/D \sim 0.4$ near the spinon
condensation-transition in Fig. 2, we pass from the
antiferromagnetic Mott insulator to the d-wave superconductor
through the spin-gapped incoherent metal with pairing
fluctuations. Compared to the high $T_c$ phase diagram at zero
temperature, the Pseudogap phase may be identified with the
NFL$\Delta$, i.e., spin-gapped incoherent metal with preformed
pair excitations in the strong coupling analysis of spin
fluctuations for the BCS-HF effective theory. The existence of
this intermediate non-Fermi liquid metal can conceptually explain
why offset of superconductivity gives rise to incoherence of
elementary excitations.\cite{ARPES,INS} In the deconfined spin-gap
phase (NFL$\Delta$) both the spin-fluctuation and electron spectra
cannot be coherent as shown in the above, and only two-particle
continuum should be observed.

In the spinon-condensed phase with
fermion pairing, condensed spinons are confined with fermions to
form electron quasiparticles (Higgs phase), thus showing the
coherent peak in the single particle spectrum. In the above
expression, when boson excitations become condensed, the
"two-particle" electron Green's function is reduced to the
original one-body Green's function, allowing the coherent peak. In
this case such coherent electron excitations also carry spin
quantum numbers. This means that the spin susceptibility is
expressed as electrons' spin correlations in the superconducting
phase. Remember that the spin susceptibility is given by the
boson-correlation function in the deconfined spin-gap phase
(NFL$\Delta$) since the spin quantum number is carried by only
boson excitations. The point is that if the resonance frequency of
spin-fluctuation modes is smaller than the superconducting gap
($2\Delta$), such resonance modes can be protected from decaying
to electron's particle-hole fluctuations, and sharply
defined.\cite{SB_INS}

The electron Green's function in the slave-boson representation
has the similar expression with that in the spin-decomposition
approach except the difference of the quantum number
assignment.\cite{RMP} Thus, boson condensation results in coherent
electron excitations in the same way as the above. On the other
hand, the spin susceptibility is expressed by fermions' spin
correlations in the slave-boson theory.\cite{SB_INS} It is
important to notice that such fermion fluctuations are not
affected by boson condensation severely in the mean-field
approximation. In particular, the fermion pairing gap, usually
called spin gap, exists in both boson condensed (superconducting)
and uncondensed (pseudogap) phases. This protects the magnetic
resonance modes from decaying, as discussed above, even in the
spin-singlet pairing phase. It may be possible to cure this result
by taking into account gauge fluctuations beyond the mean-field
approximation. However, at least in the saddle-point
approximation, the disappearance of resonance modes in the
spin-gap phase is not captured well in the slave-boson theory. One
can propose that the superconducting gap closes in the pseudogap
phase. In this case the disappearance of such resonance modes can
be explained. However, such a proposal has difficulty in
explaining the origin of pseudogap. Another mechanism for the
pseudogap should be considered in this case.

\section{Beyond the mean-field approximation}

The spin-decomposition approach has the similar spirit with the
spin-fluctuation theory\cite{Spin_fluctuation_theory} conceptually,
because spin fluctuations are taken into account more elaborately.
It is also a gauge theory with the same mathematical structure as
the slave-boson\cite{RMP} and slave-fermion
theories\cite{Shankar,Reviews,Slave_Fermion}. However, the
spin-fluctuation approach is difficult to allow the incoherent
metallic phase, which corresponds to a new stable fixed point
different from the Fermi liquid phase in the renormalization group
sense. A weak coupling approach such as the spin-fluctuation
theory is believed to allow either Landau Fermi liquid or
conventional symmetry-breaking phases.\cite{Shankar_RG} On the
other hand, the effective gauge theory [Eq. (14)], obtained in the
strong coupling approach, exhibits an infrared stable fixed point
in the presence of gauge interactions (at least without pairing
fluctuations),\cite{IR_FP} identified as the spin-gapped
incoherent metallic phase in the presence of hole doping.

It is necessary to discuss the existence of such a fixed point in
more detail, comparing with the slave-boson and slave-fermion
contexts. In the slave-boson context the spin liquid insulating
phase at half filling where bosonic charge fluctuations are gapped
is identified with such a fixed point. Integrating out gapped
boson (charge) excitations, one will obtain an effective
fermion-gauge action. When the uniform "internal" gauge flux is
considered without pairing excitations at half filling,
dissipative gauge fluctuations arise to mediate interactions
between fermion excitations, thus a non-"relativistic" gauge
theory is obtained.\cite{NFL} If the staggered flux ansatz to
allow pairing fluctuations is taken into account at half filling
in the SU(2) formulation context,\cite{RMP} the "relativistic"
QED$_3$ will be obtained, and no damping effects appear in gauge
fluctuations. Generically, the non-relativistic fermion-gauge
theory can be obtained away from half filling, since hole doping
shifts the chemical potential, making a Fermi surface.\cite{NFL}
On the other hand, the spin-gapped metal where bosonic spin
excitations are gapped corresponds to this fixed point in the
slave-fermion framework. Integrating out gapped boson (spin)
excitations, one would always obtain the non-relativistic
fermion-gauge action with damped gauge fluctuations away from half
filling, since there are no fermion excitations at half filling in
the slave-fermion approach of the t-J model. In this respect the
effective fermion-gauge theory is generically non-relativistic
with dissipative gauge fluctuations for both the slave-boson and
slave-fermion frameworks. However, there exists an important
difference in the physical point of view; the fermion excitations
carry charge quantum numbers in the slave-fermion approach while
those do spin quantum numbers in the slave-boson description.

The present gauge theoretic description has exactly the same
structure as the slave-particle theoretical framework if d-wave
pairing excitations are not taken into account. For the time
being, we consider the spin-gapped incoherent metal without
pairing fluctuations. Integrating out gapped bosonic
spin-fluctuations in Eq. (14) without the pairing term, we also
find the non-relativistic fermion-gauge action, basically the same
as the effective gauge theory of the slave-fermion description in
both the physical and mathematical points of view.

The non-relativistic fermion-gauge theory has been argued to have
an infrared stable fixed point,\cite{IR_FP} where the fixed point
value of the internal gauge coupling constant is proportional to
$1/\sigma_{\psi}$ with the fermion conductivity
$\sigma_{\psi}$.\cite{Deconfinement_Nonrelativistic1,
Deconfinement_Nonrelativistic2} This is quite reasonable since the
fixed point can arise from screening of the internal gauge charge
via fermion excitations, and the screening is associated with the
fermion conductivity. In the relativistic gauge theory the fixed
point charge is proportional to $1/N$, where $N$ is the fermion
flavor number participating in screening of gauge
interactions.\cite{Deconfinement_Relativistic_Boson,
Deconfinement_Relativistic_Fermion} In this respect the fermion
flavor number $N$ in the relativistic theory is analogous to the
fermion conductivity in the non-relativistic
theory.\cite{Deconfinement_Nonrelativistic1,
Deconfinement_Nonrelativistic2}

An important notorious question is the stability of such an
interacting fixed point against instanton excitations which result
from compactness of gauge fields.\cite{Polyakov,Fradkin,NaLee}
Although the conclusion is far from consensus, it seems to be
possible that when the fermion flavor number or conductivity is
large enough to screen the internal gauge charge, instanton
excitations can be suppressed, and the interacting fixed point
would be stable against
confinement.\cite{Deconfinement_Nonrelativistic1,
Deconfinement_Relativistic_Boson,
Deconfinement_Relativistic_Fermion,Deconfinement_Nonrelativistic2}
Recently, it was argued that the scaling dimension of an instanton
insertion operator is proportional to the fermion flavor number
$N$ at the conformal invariant fixed point of the relativistic
fermion-gauge theory. This means that instanton excitations can be
irrelevant in the large $N$ limit, expressing the stability of
such a fixed point against
confinement.\cite{Deconfinement_Relativistic_Fermion} Following
the similar strategy, one of the present authors critically
reinvestigated the stability of the interacting fixed point in the
non-relativistic fermion-gauge
theory.\cite{Deconfinement_Nonrelativistic1} Since the fermion
conductivity in the non-relativistic theory plays the similar role
as the flavor number in the relativistic one as mentioned above,
it was found that the scaling dimension of the instanton operator
is proportional to the fermion conductivity. This implies that
instanton excitations would be irrelevant at least in the large
conductivity limit corresponding to a good metal. Although we
cannot claim the appearance of deconfinement definitely, such an
anomalous spin-gapped metal may arise in principle.

Precisely speaking, the interacting fixed point associated with
the spin-gapped incoherent metal (NFL) is described by the $z = 3$
critical field theory owing to the Landau damping term that
results from gapless fermion excitations, where $z$ is the
dynamical critical
exponent.\cite{Kondo_Kim_Kim,Deconfinement_Nonrelativistic1,Chubukov_NFL}
The effective field theory is well known to cause non-Fermi liquid
physics due to scattering with massless gauge fluctuations. The
imaginary part of the fermion self-energy is given by
$\omega^{2/3}$ at the Fermi surface, implying that its real part
also has the same frequency dependence via the Kramer's Kronig
relation, thus giving rise to a non-Fermi liquid
behavior.\cite{Chubukov_NFL} Accordingly, the dc conductivity is
proportional to $T^{-5/3}$ in three dimensions and $T^{-4/3}$ in
two dimensions.\cite{NFL,Ioffe-Larkin} The coefficient $\gamma$ of
the specific heat is proportional to $- \ln T$ in three spatial
dimensions and $T^{-1/3}$ in two dimensions.\cite{Kondo_Kim_Kim}

However, there are pairing correlations in NFL$\Delta$. Such
pairing fluctuations are expected to be long-range-correlated in
space but short-range-correlated in time owing to the presence of
the Landau damping term in gauge fluctuations. In the presence of
pairing fluctuations we don't know how such fluctuations modify
the fixed point of NFL without pairing excitations. Since the
pair-pair correlation function is expected to be
singular,\cite{Kirkpatrick} these pairing excitations will modify
the gauge dynamics, which may change the dynamical critical
exponent. Accordingly, this will modify the transport and
thermodynamics. The role of pairing fluctuations in the NFL
physics needs further investigation.

\section{Discussion and summary}

The emergence of the non-Fermi liquid phase in the doped
antiferromagnetic Mott insulator can also be supported by its one
dimensional analogue. It is well known that low energy physics of
the undoped quantum spin chain can be described by the O(3)
non-linear $\sigma$ model with Berry phase.\cite{AF_1D} Utilizing
the CP$^1$ representation, one can express the non-linear $\sigma$
model in terms of bosonic spinon excitations interacting via
compact U(1) gauge fluctuations in the presence of the Berry phase
contribution. Since the Berry phase term is ignorable in the case
of integer spin, strong quantum fluctuations originating from low
dimensionality lead the integer spin chain to be disordered,
causing spinon excitations gapped.\cite{AF_1D} Such fractionalized
excitations are confined via strong gauge fluctuations, resulting
in spin excitons (particle-antiparticle bound states) as
elementary excitations. In the case of half-odd integer spin the
Berry phase plays a crucial role to cause destructive interference
between quantum fluctuations, thus weakening spin fluctuations.
Owing to the Berry phase contribution the half-odd integer spin
chain is expected to be ordered. But, low dimensionality leads the
system to be not ordered but critical, causing the spin-boson
excitations gapless.\cite{AF_1D} These spinon excitations are
deconfined because their critical fluctuations weaken gauge
interactions via screening.

When holes are doped into the antiferromagnetic spin chain,
Shankar showed that doped holes can be expressed by massless Dirac
fermions and these charge-fermions interact with the spin-bosons
via U(1) gauge fluctuations.\cite{Shankar} The presence of
massless Dirac fermions alters the resulting phase completely.
Massless Dirac fermions are well known to kill the Berry phase
contribution in the bosonization framework.\cite{Witten} Then, the
spinon excitations in the doped half-odd integer spin chain are
expected to be massive like those in the undoped integer spin
chain. But, these spinons are not confined because gauge
fluctuations become massive due to the presence of massless Dirac
fermions, thus ignored in the low energy
limit.\cite{Shankar,Witten} In the bosonization framework massless
Dirac fermions exhibit superconducting correlations. As a result,
the doped antiferromagnetic spin chain is identified with a
spin-gapped superconducting phase. Although the mechanism of
deconfinement in the U(1) spin-decomposition gauge theory is
completely different from that of the effective theory for the
doped spin chain, the spin-gapped incoherent metal with preformed
pairing of the CP$^{1}$ gauge theory is quite analogous to the
spin-gapped superconducting state, thus regarded as the high
dimensional realization of one dimensional deconfined spin-gapped
phase.

The present study is motivated by the possible existence of an
anomalous spin-gapped metal in the slave-fermion approach of the
t-J model. Such a non-Fermi liquid state was argued to be
analogous to the spin liquid Mott insulating phase in the
slave-boson approach of the t-J model. Although the spin-gapped
incoherent metal in the slave-fermion theory is quite appealing,
we discussed that it is difficult to incorporate superconducting
correlations into the slave-fermion framework due to the
fermionic statistics of charge degrees of freedom and the absence
of pairing interactions between charge fluctuations in the naive
mean-field scheme. In this paper we have developed how to
introduce d-wave superconductivity, keeping the slave-fermion
scheme.

It is also an important question of this paper how the
conventional theoretical framework such as the BCS-HF scheme can
give rise to the anomalous metallic phase of the slave-fermion
theory. In this paper we found a possible connection between the
slave-fermion approach and BCS-HF scheme (beyond), showing how the
non-Fermi liquid metal arises from the BCS-HF framework.
The spin-fluctuation approach was the first
candidate, but it was not adopted in this paper because such a
Fermi-liquid based weak-coupling approach is difficult to allow
the stable non-Fermi liquid phase beyond quantum criticality in
the view of its theoretical structure. Instead, we applied the
CP$^1$ decomposition scheme as the strong coupling framework.
Performing the Hubbard-Stratonovich transformation and appropriate
saddle-point approximation, we found an effective gauge theory,
quite parallel to the slave-fermion gauge theory. The
present CP$^1$ gauge theory allows pairing fluctuations between
charge degrees of freedom.

Performing the mean-field analysis, we found the phase diagram of
the effective gauge theoretical framework. Effects of spin
fluctuations strongly modified the BCS-HF phase diagram [Fig. 1],
resulting in the phase diagram of Fig. 2. In particular, the
spin-gapped incoherent metallic phase is found when Hubbard-U
interactions are beyond a certain critical value. Such a non-Fermi
liquid metal is certainly expected since it corresponds to that of
the slave-fermion theoretical framework. This non-Fermi liquid
phase is modified due to pairing correlations of charge degrees of
freedom. Actually, we found preformed pairing excitations in the
non-Fermi liquid metal near the d-wave superconducting phase.

We have also discussed the stability of such a mean-field phase
beyond the mean-field approximation, allowing gauge fluctuations.
We claimed that low energy physics of the spin-gapped incoherent
metal is described by the non-relativistic fermion-gauge
Lagrangian with damped gauge fluctuations, and such an effective
field theory gives rise to an infrared stable fixed point. This
interacting fixed point identifies the non-Fermi liquid metal
beyond the mean-field description. We discussed the stability of
such a fixed point against instanton excitations, and argued that
the fixed point can be stable against confinement when the fermion
conductivity is sufficiently large.

\begin{center}
{\bf ACKNOWLEDGEMENTS}
\end{center}

K.-S. Kim thanks K. Park for enlightening discussions.


\begin{thebibliography}{9}
\bibitem{Comment_SF} Performing the saddle-point analysis in the
slave-fermion approach of the t-J model, one finds that it is not
easy to obtain such a spin-gapped state at zero temperature, where
boson excitations are gapped. Actually, considering the uniform
flux phase of $\chi_{ij}^{\psi} = \chi_{\psi}$ and $\chi_{ij}^{b}
= \chi_{b}$ in the slave-fermion Lagrangian of Eq. (4), only a
boson-condensed phase is found at zero temperature even far away
from half filling. One may modify this artificial ordering of the
mean-field analysis far away from half filling, introducing
frustrating interactions such as next-nearest-neighbor or
ring-exchange interactions, or competing orders such as the
$\pi$-flux phase or superconducting state if possible. In
particular, we are investigating the role of $\pi$-flux in the
boson condensation, allowing next-nearest-neighbor exchange
interactions.\cite{Kim_Jia} We expect that effects of both
frustration and competition may give rise to the condensation
transition in the mean-field analysis of the boson sector. On the
other hand, such boson excitations can be gapped at finite
temperatures owing to thermal fluctuations, thus finite
temperature behaviors may be governed by the spin-gapped metallic
physics.\cite{Slave_Fermion}
\bibitem{Gauge_Transformation} The effective Lagrangian will be
invariant under the U(1) gauge transformation \bqa &&
\psi_{i\sigma} \rightarrow e^{i\sigma\theta_{i}}\psi_{i\sigma} ,
~~~~~ z_{i\sigma} \rightarrow e^{ - i\theta_{i}}z_{i\sigma} ,
\nonumber \eqa if the following link variables are transformed as
\bqa && c_{ij} \rightarrow c_{ij} - (\theta_{i} - \theta_{j}) ,
~~~~~ \vartheta_{ij}^{\psi} \rightarrow \vartheta_{ij}^{\psi} -
(\theta_{i} - \theta_{j}) . \nonumber \eqa
\bibitem{AFdSC_dSC} Considering the case when bosonic spinons are
condensed, one should introduce the condensation amplitude $Z =
|\langle{z}_{i\sigma}\rangle|^{2}$ in the self-consistent
equations for the boson sector. Solving these modified equations,
one would find such a transition line between the AF$+$dSC and
dSC. In this paper we do not perform such an analysis because our
objective is to show the existence of the spin-gapped non-Fermi
liquid phase. Starting from this solution of the present
self-consistent equations [Eqs. (16) and (17)], we can find
several instability lines where such an anomalous metal begins to
be unstable. This analysis results in the phase diagram of Fig. 2.
\bibitem{Ioffe-Larkin} One may expect that the response function
such as the conductivity is given by the Ioffe-Larkin combination
rule [L. B. Ioffe and A. I. Larkin, Phys. Rev. B {\bf 39}, 8988
(1989)]. However, this is not true owing to the presence of
$\sigma = \pm$ in the gauge coupling. Considering the conductivity
for example, one finds that it is solely given by the fermion
conductivity. Expanding the effective action to the gaussian order
for the U(1) gauge field $c_{\mu}$, the partition function is
given in a highly schematic form \bqa && Z = \int{Dc}\exp\Bigl[-
\frac{1}{2}\frac{\delta^{2}S_{z}}{\delta{c}^{2}}{c}^{2} -
\frac{1}{2}\Bigl(\frac{\delta^{2}S_{\psi}}{\delta{c}^{2}}{c}^{2} +
2\frac{\delta^{2}S_{\psi}}{\delta{c}\delta{A}}{c}{A}  \nn && +
\frac{\delta^{2}S_{\psi}}{\delta{A}^{2}}{A}^{2} \Bigr) \Bigr] =
Z_{0}\exp\Bigl[ -
\frac{1}{2}\frac{\delta^{2}S_{\psi}}{\delta{A}^{2}}{A}^{2} +
\frac{1}{2}\frac{\frac{\delta^{2}S_{\psi}}{\delta{a}\delta{A}}}
{\frac{\delta^{2}S_{z}}{\delta{c}^{2}}+\frac{\delta^{2}S_{\psi}}{\delta{c}^{2}}}A^{2}\Bigr]
, \nonumber \eqa where $S_{z}$ ($S_{\psi}$) is the boson (fermion)
action. Differentiating the above effective action by an
electromagnetic field $A_{\mu}$ twice, one can obtain the
conductivity expression. The presence of $\sigma = \pm$ results in
${\delta^{2}S_{\psi}}/{\delta{c}\delta{A}} = 0$ because
correlations between the charge current and spin current are zero.
As a result, the conductivity is given by the fermion contribution
only.
\end{thebibliography}
\end{document}